\documentstyle[12pt,fullpage,psfig]{article}
\begin{document}
\begin{center}
\hfill \parbox{1.5in}{hep-lat/9805023\\MIT-CTP-2738\\ITP-SB-98-38}\\
\vskip 1in

{\LARGE\bf 
Critical region of the finite temperature chiral transition
}

\vskip 1in

J.B. Kogut$^{\rm a,b}$, M.A. Stephanov$^{\rm c}$, and
C.G. Strouthos$^{\rm a}$

\vskip 1in

$^{\rm a}$ {\it
Department of Physics, University of Illinois at Urbana-Champaign, 
Urbana 61801-3080
}\\
$^{\rm b}$ {\it
Center for Theoretical Physics, Laboratory for Nuclear Science and
Department of Physics, MIT, Cambridge, MA 02139
}\\
$^{\rm c}$ {\it 
Institute for Theoretical Physics, SUNY, Stony Brook, NY 11794-3840
}

\vskip 1.5in

{\large\bf Abstract}
\bigskip
\end{center}

We study a Yukawa theory with spontaneous chiral symmetry breaking
and with a large number $N$ of fermions near
the finite temperature phase transition.  Critical properties in such
a system can be described by the mean field theory very close to the
transition point. We show that the width of the region where
non-trivial critical behavior sets in is suppressed by a certain power
of $1/N$. Our Monte Carlo simulations confirm these analytical results.
We discuss implications for the chiral phase transition in QCD.

\newpage

\section{Introduction}

The transition in QCD separating the high temperature quark-gluon plasma
phase from the low temperature hadronic phase has been studied
intensively in the last decade.
Understanding the properties of this transition is becoming
increasingly important in view of recent experimental progress
in the physics of ultrarelativistic heavy ion collisions.

Since the $u$ and $d$ quark masses are small the dynamics of the
finite temperature transition
is affected by the phenomenon of chiral symmetry restoration
which occurs in the limit when the quark masses are put to zero.
In this limit QCD with two massless quarks has a global
SU(2)$_L\times$SU(2)$_R$ symmetry which is spontaneously broken to 
SU(2)$_V$ at low temperatures. It can be argued that if the restoration
of this spontaneously broken symmetry proceeds through a second-order phase
transition the critical properties of this transition are
in the universality class of the classical O(4) spin model in three
dimensions \cite{PiWi84,RaWi93}. This means that the leading singular behavior
of thermodynamic quantities can be predicted, i.e., it is
given by universal O(4) critical exponents. However, universality
does not answer more detailed questions such as how this criticality
is approached and what is the width of the region of parameters
in which this singular behavior sets in. These questions require
more detailed knowledge of the dynamics of the theory.

In this paper we discuss a phenomenon which 
is related to the way the critical behavior sets in for a certain
class of theories with second-order chiral symmetry restoration transition
at finite temperature, $T_c$. These are theories with a large number
of fermion species $N$. On general grounds, one could expect that the
universal critical, or scaling, behavior near $T_c$ sets in as soon as
the correlation length, $\xi$, of the fluctuations of the chiral
condensate exceeds $1/T_c$. However, as we show in this paper,
if the number of fermions involved in the chiral symmetry breaking
is large, the critical behavior which sets in when $\xi$ exceeds $1/T_c$
is given by the mean field theory, rather than by the arguments
based on dimensional reduction and universality as in \cite{PiWi84}.
The critical behavior given by these latter arguments
sets in much later, closer to $T_c$,
when the correlation length $\xi$ exceeds $N^x/T_c$. 
Below we shall determine the value of the positive exponent $x$.

The phenomenon of the non-trivial critical region suppression 
has been observed in the Yukawa  model \cite{RoSp94} and in the Gross-Neveu%
\footnote{In this paper we shall use the terms ``Gross-Neveu model''
and ``Nambu-Jona-Lasinio model'' interchangeably, especially
when it concerns a theory with a four-fermion interaction in 2+1 dimensions.}
model \cite{KoKo94}
using large-$N$ expansion and was confirmed by lattice Monte Carlo
calculations in \cite{KoKo94}. These large-$N$ results predicting
mean-field critical behavior were in an apparent contradiction
with the general arguments of \cite{PiWi84}. In this paper we show
that both critical regimes are realized in the vicinity
of $T_c$, but in separate scaling windows, one following the other.

\section{Large $N$ Yukawa theory near $T_c$}

In this section we consider a general Yukawa theory in $d$ dimensions,
$2<d\le4$, with a large number $N$ of 
fermion species and at finite temperature $T$.
As argued in \cite{HaHa91}, in the continuum limit both Yukawa model 
and Gross-Neveu, or Nambu-Jona-Lasinio, models with a four-fermion
interaction define the same theory.
In the absence of a bare fermion mass there is a (chiral) symmetry, 
which can be broken spontaneously
at low temperature with a suitable choice of couplings. 
This symmetry is restored at some finite temperature $T_c$.
We are interested in the nature of this phase transition.

In the absence of the fermions (or when the Yukawa coupling is zero) the
nature of the transition is rather well known. It depends on the symmetry
of the model which is restored at $T_c$. 
In this paper, for simplicity, we consider
a model with $Z_2$ symmetry. Similar results will also 
apply to theories with other symmetry groups (e.g., SU(2)$\times$SU(2),
as in QCD with 2 massless quarks), as long as
the temperature driven symmetry restoration transition is of 
the second order.

One can argue that the critical behavior of a quantum theory of a scalar field 
near $T_c$ is the same as in a classical scalar theory with the same
symmetry. The argument is based on two expected properties of the
model: dimensional reduction and universality.  In the Euclidean
formulation the quantum scalar field is defined in a $d$-dimensional box
with the extent in the imaginary (Matsubara) time dimension equal
to $1/T$. When the diverging correlation length, $\xi$, becomes much
larger than $1/T_c$ long wavelength fluctuations of the
field become effectively $(d-1)$-dimensional, i.e., on their scale the box
looks like a $(d-1)$-dimensional ``pancake''. Since such fluctuations
determine the critical behavior in the model one can expect
that, by universality, the critical exponents are the same as in
a $d-1$-dimensional theory with the same symmetry. This $d-1$-dimensional
theory is obviously a classical field theory at finite temperature.
One can also understand this realizing that the classical thermal
fluctuations whose energy is $O(k_BT)$
dominate over the quantum fluctuations with energy $O(\hbar\omega)$ for
soft modes of the field. 

Another common way of describing this phenomenon in perturbation
theory is to consider
Fourier decomposition of the field into discrete Matsubara frequency
components. Non-zero frequency acts as a mass term of order $\pi T$
for the $(d-1)$-dimensional components of the field. Near $T_c$ this mass is
much larger than the mass of a component with zero Matsubara
frequency.  The dimensional reduction is
then equivalent to the decoupling of the modes with nonzero frequencies.

We want to understand what happens in this theory near $T_c$ when one
turns on the Yukawa coupling. Dimensional reduction and universality
arguments suggest that the critical behavior of the theory should not
change.  This is based on the observation that there is no zero
Matsubara frequency for the fermion fields due to antiperiodic
boundary condition in the Euclidean time. Therefore all fermion modes
should decouple at $T_c$.  In other words, the fermion fields do not
have a classical limit and do not survive quantum-to-classical
reduction at $T_c$ \cite{St95}.

However, as was demonstrated in \cite{KoKo94}, the 
fermions do affect the behavior near $T_c$ in a certain way. We
shall show that this
happens when there are ``too many'' of them. The theory can be solved
in the limit when the number of fermions, $N$, is large. In this
limit the theory has mean-field critical behavior near $T_c$
\cite{KoKo94,RoSp94}. This is different from the critical behavior of
the corresponding classical scalar field theory. The mean-field
behavior was also observed in numerical Monte Carlo calculations near
$T_c$ \cite{KoKo94}.

Here we show that such mean-field behavior can be reconciled with
the standard arguments of dimensional reduction and universality.
The phenomenon which leads to an apparent contradiction is the
suppression of the width of the non-mean-field critical region by a power 
of $1/N$.

We consider the following model with one-component scalar
field and $Z_2$ symmetry in $d$ dimensional Euclidean
space:
\begin{equation}\label{yukawa}
{\cal L} = {1\over2}(\partial\phi)^2 + {1\over2}\mu^2\phi^2
+ \lambda\phi^4  + \sum_{f=1}^{N}\bar\psi_f \left( 
\partial\hskip -.5em / + {g\phi}
\right) \psi_f
\end{equation}

We regularize the model by some momentum cutoff, $\Lambda$. The
cutoff can be removed if $d<4$ but we shall keep it finite
to compare with a corresponding lattice theory. There are two other
important scales in the theory: the temperature, $T$, and
the physical mass, $m$, which we identify with the mass
of thermal excitations of the scalar field. 
Near $T_c$ this mass, $m$, is significantly
different from the zero temperature mass $m_0$, since 
$m$ vanishes at the critical temperature.
It is the mass $m$, or the correlation length $1/m$,
which is important for the critical behavior.
The mass $m$ is a natural measure (more natural than, say,
$T-T_c$) of the distance from the criticality.%
\footnote{The whole theory of critical scaling is based on this observation.
We shall also use this fact more explicitly when we consider 
a lattice theory.}
Therefore, near the finite temperature phase transition we have 
the following hierarchy of scales: $\Lambda\gg T \gg m$.

Let us consider the renormalization group (RG) evolution of the
couplings from the scale of $\Lambda$ down to the scale of
$T$ and then from $T$ down to $m$.
We focus on the quartic
self coupling of the scalar field, $\lambda$. The evolution from
the scale $\Lambda$ to the scale $T$ is governed by the
RG equations of the $d$-dimensional quantum Yukawa model. 
After that, at the scale of
$T$, we pass through a crossover region due to the fact
that the fermions and nonzero Matsubara frequencies of the scalar fields
do not contribute to the evolution below $T$ (the decoupling). The evolution
below $T$ is governed by the RG equations of the scalar $\phi^4$
theory in $d-1$ dimensions.

If the window of scales between $\Lambda$ and $T$ is wide enough (as
it is in the continuum limit $\Lambda\to\infty$) the
value of the renormalized coupling $\lambda$ at the scale $T$,
$\lambda(T)$, is close to the infrared fixed point of the
$d$-dimensional Yukawa theory. In the large-$N$ limit one can
calculate this value \cite{Zi96}:
\begin{equation}\label{lambda}
\lambda(T) \sim {(4-d)\ T^{4-d}\over N} \quad\mbox{for} \quad 2<d<4.
\end{equation}
The case $d=4$ is special. The infrared fixed point is trivial
and is approached logarithmically as $\Lambda/T\to\infty$:
\begin{equation}\label{lambda,d=4}
\lambda(T) \sim {1\over N\ln(\Lambda/T)} \quad\mbox{for}\quad d=4.
\end{equation}
This value provides the starting point, 
$\lambda_{d-1}=T\lambda(T)$,
for the evolution of this coupling below $T$ in the $\phi^4$ theory in
$d-1$ dimensions. For large $N$ this coupling is small. 
As we shall see shortly, this is the reason why the critical region 
where one observes 
non-trivial critical behavior is reached only very close to the
phase transition.

The phenomenon of the suppression of the width of a non-trivial
critical region is common in condensed matter physics \cite{PfTo77}.
BCS superconductors provide the most well-known example
of a system where criticality near $T_c$ is described
by the Landau-Ginzburg mean-field theory. The width, $\Delta T$,
of the region around $T_c$ where the mean-field description
breaks down is tiny.
There are systems where the width of the non-trivial
critical region is small but measurable. In such systems
one can observe the crossover between a mean-field and
a non-trivial scaling.

The quantitative relation between the size of the non-trivial, 
non-mean-field  critical region, the Ginzburg region, and certain
parameters of a given system is known as the  Ginzburg criterion.
In superconductors such a parameter is a small ratio $T/E_F$,
i.e., the width of the Ginzburg region is suppressed by a power
of this parameter. In this paper, we show that in a field
theory with large number of fermions, such as (\ref{yukawa}),
the width of the Ginzburg region is suppressed by a power of $1/N$.

The Ginzburg criterion can be obtained by estimating
the effects of fluctuations within the mean-field
approximation. When the fluctuations become large
the mean-field approximation
breaks down because of self-inconsistency.
A measure of the importance of the fluctuations, or the size
the corrections to the mean-field, is the value
of the effective selfcoupling of the scalar field, $\lambda$.
If the coupling $\lambda$ is small the effect of 
the fluctuations is also small and the theory can
be described by the mean-field approximation very well. However, for
the $\phi^4$ theories in less than four dimensions%
\footnote{Here again the case of four dimensions is special. However,
we are now discussing theories in $d-1$ dimensions and $1<d-1\le3$.}
fluctuations always become important close enough to the phase transition.
This happens roughly when the coupling $\lambda_{d-1}$ on the
scale of $m$ is not small anymore. Since $\lambda_{d-1}$
has nonzero dimension equal to $5-d$, it should be compared to
$m^{5-d}$. In this way, and with the help of (\ref{lambda}),
one arrives at the following Ginzburg criterion for
the applicability of the mean-field scaling:
\begin{equation}
m \gg {T\over N^x} \ , \quad x={1\over5-d} \ .
\label{ginzburg}
\end{equation}
In the special case of $d=4$, using (\ref{lambda,d=4}) one finds:
\begin{equation}
m \gg {T\over N\ln(\Lambda/T)} \quad \mbox{for} \quad d=4\ .
\label{ginzburg,d=4}
\end{equation}

\begin{figure}[hbt]
                \centerline{ \psfig{silent=,file=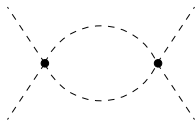,height=0.6in} }
\caption[]{An example of a graph whose contribution
breaks the mean field approximation and the large-$N$ expansion near $T_c$.}
\label{fig:graph}
\end{figure}

Alternatively, one can compare the size of the one-loop correction
in the effective $\phi^4$ theory in $d-1$ dimensions to the 
bare $\lambda_{d-1}$.
The loop correction becomes important when (\ref{ginzburg}) is
violated.%
\footnote{In fact, the criterion (\ref{ginzburg,d=4}) is well-known 
in the form $\lambda T_c/m \ll 1$ \cite{We74} as the
criterion for the applicability of perturbation theory near a finite
temperature phase transition.}
Indeed, consider the contribution, $\Delta \lambda$
of a graph such as in Fig.\ref{fig:graph} to the effective
quartic coupling $\lambda$. This contribution diverges
when $m\to0$:
\begin{equation}
\Delta \lambda \sim {T[\lambda(T)]^2\over m^{5-d}},
\end{equation}
where we integrated over fluctuations
in the window between $T$ and $m$ with $m\ll T$.
Thus, this one-loop contribution, $\Delta \lambda$,
is negligible compared to $\lambda(T)$
if $T\lambda(T)\ll m^{5-d}$, which together with (\ref{lambda})
leads to (\ref{ginzburg}). Note also that, since
$\lambda(T)\sim 1/N$, the contribution of the graph in 
Fig.\ref{fig:graph} should be subleading in the large-$N$ expansion.
Therefore, the large-$N$ expansion breaks down when (\ref{ginzburg})
is violated.

The Ginzburg criterion tells us that for masses $m$ inside the window
$T \gg m \gg T/N^x$ the mean-field scaling holds,
while for smaller masses $m \ll T/N^x$ (i.e., closer to the transition) 
the non-trivial $d-1$ Ising scaling sets in. We see that the size
of this latter, non-trivial critical region is suppressed at large
$N$.%
\footnote{Note that there is no proportionality constant in the
Ginzburg criterion (\ref{ginzburg}). This constant would depend
on the definition of the boundary of the mean-field region
which is naturally ambiguous. The Ginzburg criterion tells
us how this boundary moves as a function of $N$.}

\section{Lattice theory}

In this section we analyze how the effect of the suppression
of the non-trivial critical region manifests itself
 in the lattice formulation
of the theory. For simplicity, we shall discuss the case $d=3$,
The generalization to arbitrary $d$, $2<d\le4$, can be done as in the
previous section.
We consider the following lattice discretization of the theory
(\ref{yukawa}) in $d=3$:
\begin{equation}\label{yukawa-lat}
S = \sum_{{\tilde{x}}}\left(
- \kappa \sum_\mu \phi_{{\tilde{x}}+\hat\mu}\phi_{\tilde{x}}
+ \lambda_{\rm lat}\phi_{\tilde{x}}^4 
+{\beta N\over4} \phi^{2}_{\tilde{x}}
\right)
+\sum_{i=1}^{N/2} \left( 
\sum_{x,y} \bar{\chi}^{i}_{x} M_{x,y} \chi^{i}_{y}
+{1\over8} \sum_{x} \bar{\chi}^{i}_{x} \chi^{i}_{x} 
\sum_{ \langle \tilde{x},x \rangle} \phi_{\tilde{x}} \right) ,
\end{equation}
where $\chi^{i}$ and $\bar{\chi}^{i}$ are  Grassmann-valued staggered
fermion fields defined on the lattice sites; the scalar field $\phi$
is defined on the dual lattice sites, and the symbol 
${ \langle \tilde{x},x \rangle} $ denotes the set of 8 (i.e., $2^d$) 
dual lattice sites $\tilde{x}$ surrounding the direct
lattice site $x$. The fermion kinetic (hopping) matrix $ M $ is given by
\begin{equation}
M_{x,y} = \frac12 \sum_{\mu} \eta_{\mu}(x)
\left[ \delta_{y,x+\hat{\mu}} - \delta_{y,x-\hat{\mu}} \right],
\end{equation}
where $\eta_{\mu}(x)$ are the Kawamoto-Smit phases 
$(-1)^{x_{1}+...+x_{\mu-1}}$.
The cubic lattice has
$L_{s}$ lattice spacings $a$ in spatial directions and $L_t$ lattice
spacings in the temporal direction.
The cutoff scale can be defined as $\Lambda=
1/a$, the temperature is given by $T=1/(L_ta)$ and the mass is $m=1/(\xi a)$,
where $\xi$ is the correlation length of the scalar field.
To reach a continuum limit one has to satisfy the following two
conditions: $\Lambda\gg T$ and $\Lambda\gg m$. The condition $\Lambda\gg T$
requires a lattice with sufficiently large $L_t\gg 1$. The parameters
of the action should then be tuned towards their critical values
where the correlation length $\xi \gg 1$. This satisfies the 
condition $\Lambda\gg m$.
 
The condition $\xi \gg 1$, or $ma=1/\xi\to0$, specifies a
2-dimensional critical surface in the space of 3 parameters $\kappa$,
$\lambda_{\rm lat}$ and $\beta$. One expects
that a continuum limit taken at any generic point of this surface
defines the same theory \cite{HaHa91,FoFr94}. 
(This is the meaning of the equivalence
between Yukawa and four-fermion theories). Therefore we can fix
two out of three bare parameters, $\kappa=\lambda_{\rm lat}=0$ and
tune a single parameter $\beta$ to criticality: $1/\xi\to0$. 

\begin{figure}[htb]
		\centerline{	\psfig{silent=,file=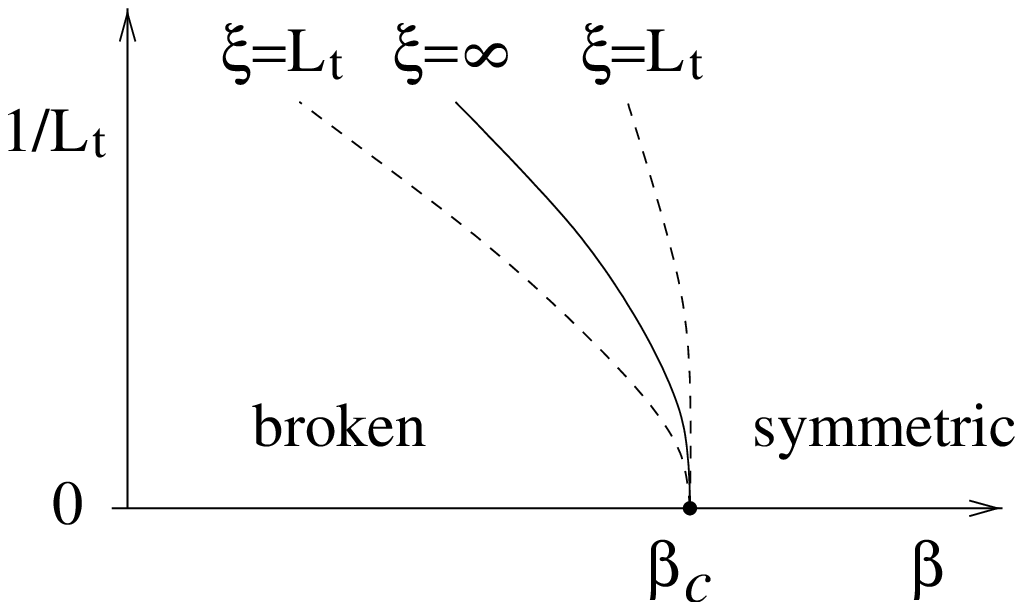,height=1.7in}
				\hspace{0.5in}
				\psfig{silent=,file=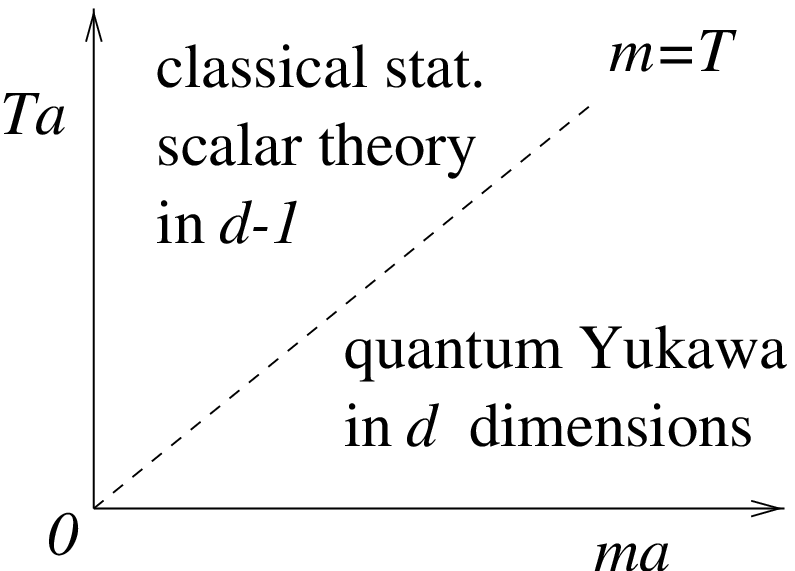,height=1.7in}}
\caption[]{A schematic phase diagram (a) of a lattice Yukawa theory in the
plane of $Ta\equiv1/L_t$ and the lattice action parameter $\beta$. 
The solid line is the phase boundary. The dashed lines show
the location of points where the correlation length, $\xi$, reaches $L_t$.
The same phase diagram (b) but the distance from the critical line is
expressed in terms of a natural variable
$ma=1/\xi$ and only one side (either symmetric or broken) is shown.
Various continuum limits correspond to approaching the origin in (b). The
slope determines the ratio $T/m$ in the resulting continuum theory.
The values of $L_t$ and thus of $Ta$ are discrete, but this is of no 
importance to our discussion.}
\label{fig:phadi1}
\end{figure}

The phase diagram as a function of $\beta$ and $Ta=1/L_t$ looks like
in Fig.~\ref{fig:phadi1}a.  A natural measure of the distance from the
criticality is $ma=1/\xi$.  Trading the lattice parameter $\beta$ for
$ma$ we obtain the phase diagram of Fig.~\ref{fig:phadi1}b.

The line $T=m$ separates the regions of quantum and classical behavior
or, equivalently, the regions of $d$-dimensional and
$(d-1)$-dimensional behavior. Below this line, when $T\ll m$, the
correlation length $\xi$ is smaller than the extent in the time
direction $L_t$ and the system behaves as a quantum Yukawa model in
2+1 dimensions. Above the line, when $T\gg m$, the correlation length
$\xi$ is much larger than $L_t$, the system looks like a ``pancake''
and behaves as a 2-dimensional classical statistical theory of a
scalar field.

\begin{figure}[htb]
		\centerline{	\psfig{silent=,file=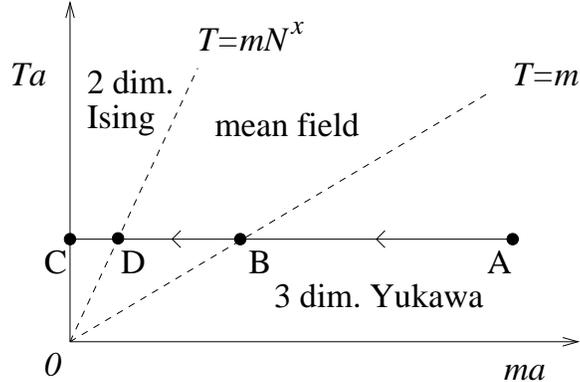,height=2in}}
\caption[]{The same phase diagram as in Fig.~\ref{fig:phadi1}b for
a Yukawa theory with large $N$. The trajectory
ABC corresponds to changing some lattice parameter to approach
criticality on a lattice with a given $L_t\equiv1/Ta$. The point
A corresponds to the correlation length $\xi\equiv1/ma\sim1$,\,\, 
$\xi(B)=L_t$,\,\,  $\xi(D)=L_tN^x$,\,\,  and $\xi(C)=\infty$.  
In $d=3$: $x=1/2$.
}
\label{fig:phadi3}
\end{figure}

Now consider changing the parameter $\beta$ on a given lattice, i.e.,
at fixed $Ta=1/L_t$, so that we move along a trajectory such as ABC on
Fig.~\ref{fig:phadi3}. The effective (long distance) coupling
$\lambda(\xi)$ follows the evolution governed by the RG equations of
the Yukawa model from A to B.\ Near the point B it reaches some value,
$\lambda(L_t)$, which, if $L_t$ is large enough, is given by the
infrared fixed point and is $O(1/N)$.  As we continue to increase
$\xi$ from B to C the coupling $\lambda(\xi)$ evolves according to the
RG equations of the 2-dimensional $\phi^4$ theory.  Since
$\lambda(L_t) \sim 1/N$, in order to reach the non-trivial critical region one
needs to go to the correlation length $\xi\sim L_tN^x$, with $x=1/2$
in $d=3$, according to the Ginzburg criterion~(\ref{ginzburg}).

Thus the line of a crossover with a slope $Ta/ma = N^x$ divides
the region, $T>m$, of a 2-dimensional, or classical, behavior
into two subregions, or windows. One window, where the mean-field
approximation works ($L_t\ll\xi\ll L_tN^x$), and another window, where
the non-trivial critical behavior sets in ($\xi\gg L_tN^x$).

\section{Monte Carlo}

We performed Monte Carlo simulations of the Gross-Neveu model in
$d=2+1$ dimensions at finite temperature to test the results of the
analysis of the previous section. We chose the $Z_{2}$ four-Fermi
model because it is relatively easy to simulate.  We used a Hybrid
Monte Carlo method described in \cite{HaKoKo93}, which proved to be
very efficient for our purposes.  Since the chiral symmetry is
discrete we were able to simulate the model directly in the chiral
limit $m=0$. This allowed particularly accurate determination of the
critical properties.  The action is that of the Yukawa lattice theory
(\ref{yukawa-lat}) with $\kappa=\lambda=0$, and we tuned $\beta$ to
reach criticality.  The long-wavelength (continuum limit) behavior of
such a theory is determined by the infrared fixed point, which is the
same~\cite{HaHa91,FoFr94} in the more general Yukawa model
(\ref{yukawa-lat}) and in the Gross-Neveu model.

We used the following two methods to optimize the performance of the
Hybrid Monte Carlo procedure. The first method consisted of tuning the
effective number of fermion flavors $N^{'}$, which is used
during the integration of the
equations of motion along a microcanonical trajectory, so as to
maximize the acceptance rate of the Monte Carlo procedure for a fixed
microcanonical time-step $d\tau$. As the lattice size was increased,
the time step $d\tau$ had to be taken smaller and the optimal $N'$
approached $N$.  For example, for an  $N=4$ theory on a $6
\times 36^{2}$ lattice the choices $d\tau=0.15$ and $N'=4.036$
gave acceptance rates greater than $95\%$ for all couplings of
interest. To maintain this acceptance rate on a $6 \times 60^{2}$
lattice we used $d\tau=0.11$ and $N'=4.016$, while on a $6 \times
80^{2}$ lattice we used $d\tau=0.10$ and $N'=4.012$.  In the
second method the Monte Carlo procedure was optimized by choosing the
trajectory length $\tau$ at random from a Poisson distribution with
the mean equal to $\bar{\tau}$. This method of optimization, which
guarantees ergodicity, was found to decrease autocorrelation times
dramatically \cite{HaKoKoRe94}. For most of our runs we used the
average trajectory length $\bar{\tau} \simeq 2.0$.  As usual, the
errors were calculated by the jackknife
blocking, which accounts for correlations in a raw data set.

As will be seen below, we used values of the lattice coupling $\beta$
sufficiently close to the critical value, $\beta_c$, so that we are close
to the continuum limit $\Lambda\gg m$, where $m$ is the thermal mass,
and the scaling behavior is not affected by lattice artifacts.
In addition, we verified that also another important physical parameter, 
the zero-temperature mass, $m_0$, is sufficiently smaller than the cutoff
$\Lambda$. We ran on lattices with large $L_t$, such as $20^3$ with
$N=4$ for $\beta=0.600 - 0.750$, and $16^3$ with $N=24$ for $\beta=0.725 - 0.875$,
and determined the value of the scaling exponent $\beta_m$.
For $N=12$ we can use the results from \cite{HaKoKo93}.
We found values in agreement with the analytical prediction
$\beta_m=1+{\cal O}(1/N^2)$ for the $T=0$ scaling \cite{HaKoKo93}. This confirms that
for our values of the coupling $\beta$ the lattice theory remains in the
scaling window for all range of temperatures down to $T=0$ and effects of the
lattice are negligible.

\subsection{Exponents from finite size scaling}

The finite size scaling (FSS) analysis is a well-established tool
for studying critical properties of phase transitions \cite{Ba83}.
The critical, singular behavior in a statistical system is caused by
the divergence of the correlation length $\xi$. 
On a finite lattice the correlation length is limited by the size
of the system and, consequently, no true criticality can be observed.
However, if the size, $L_s$, of the lattice is large, a qualitative
change in the behavior of the system occurs when $\xi\sim L_s$.
For $1\ll\xi\ll L_s$ the behavior of the system is almost the same as in
the bulk ($L_s=\infty$). However, when $\xi\sim L_s$ the behavior of the
system reflects the size and the shape of the box to which 
it is confined.
The dependence of a given thermodynamic observable, $P$, on the
size of the box, $L_s$, is singular and, according to the FSS
hypothesis, is given by:
\begin{equation}\label{fssX}
P(t,L_s) = L_s^{\rho_P/\nu}Q_P(tL_s^{1/\nu}),
\end{equation}
where $t$ is the distance from the critical point: 
$t=(\beta_c-\beta)/\beta_c$;\ \,
$\nu$ is the standard exponent of the correlation length:
$\xi\sim t^{-\nu}$; and $Q_P$ is a scaling function, which is not
singular at zero argument. 
The exponent $\rho_P$ is the standard critical exponent
for the quantity $P$: $P\sim t^{-\rho_P}$.
Studying the dependence on the size of the box, $L_s$,
and using (\ref{fssX}) one can determine such exponents.

We simulated the model with $N=12$ fermion flavors at  $\beta$
close to the critical coupling $\beta_c$. The lattice sizes ranged
from $L_{s}=12$ to $40$ for $L_{t}=4$, and $L_{s}=18$ to $50$ for
$L_{t}=6$. Periodic boundary conditions in the spatial directions were
used.  Details of the $L_{t}=6$ runs are listed in
Table~\ref{tfss}.  To perform our study most effectively we used the
histogram reweighting method \cite{FeSwe88} which enables us to
calculate the observables in a region of $\beta$ around the simulation
coupling $\beta_{\rm sim}$.

\subsubsection{Exponent $\nu$}

\begin{figure}[htb]

                    \centerline{    \psfig{silent=,file=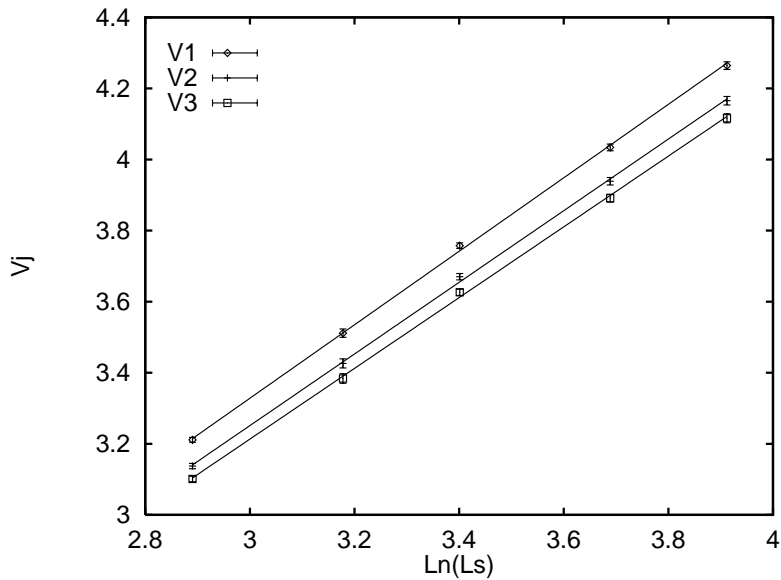,height=3in} }
\caption[]{Finite size dependence of $V_j$ at $\beta=0.7762$ for $L_{t}=6$.
All three lines have almost equal slopes.}
\label{fig:vj1}
\end{figure}

We used the following thermodynamic observables \cite{ChFeLa93} 
to determine the values of $\nu$ and the critical coupling $\beta_c$:
\begin{equation}\label{vj1} 
V_{1} \equiv 4[\phi^{3}]-3[\phi^{4}]\,,\quad
V_{2} \equiv 2[\phi^{2}]-[\phi^{4}]\,, \quad
V_{3} \equiv 3[\phi^{2}]-2[\phi^{3}]\,, 
\end{equation}
where
\begin{equation}\label{vj2} 
 [\phi^{n}]\equiv 
\ln\frac{\partial \langle \phi^{n} \rangle}{\partial \beta}\ .
\end{equation} 
One can easily find that
\begin{equation} 
V_{j}\approx(1/\nu)\ln L_{s}+{\cal V}_{j}(tL_{s}^{1/\nu}),
\end{equation}
for $j=1,2,3$. At $\beta_{c}$, i.e., $t=0$, the last term in
the r.h.s, ${\cal V}_{j}(0)$, is a constant independent of $L_s$.
Scanning over a range of $\beta$'s and looking for the value of $\beta$
at which the slope of $V_j$ versus $\ln L_s$ is $j$-independent,
as it is in Fig.~\ref{fig:vj1},
we found: $ \nu=1.00(3) $ and $ \beta_{c}=0.7762(15)$ for $L_{t}=6$, 
and $ \nu=1.00(2) $ and $ \beta_{c}=0.682(2)$ for $L_{t}=4$. 

These values of
$ \nu$ are in a very good agreement with the two-dimensional Ising 
value $\nu=1$. This confirms that the
behavior of the system sufficiently close to criticality is
non-trivial (in the mean field theory: $\nu=1/2$) 
and is given by the arguments of dimensional reduction and universality.

\subsubsection{Exponents $\beta_m$ and $\gamma$}

In this subsection we consider the following two exponents:
The exponent $\beta_m$ of the order parameter: 
$\Sigma\equiv\langle\phi\rangle\sim t^{\beta_m}$,
and the exponent $\gamma$ of the susceptibility:
$\chi\sim t^{-\gamma}$.
According to eq.~(\ref{fssX}), at the critical point, $t=0$,
the order parameter, $\Sigma$, and its susceptibility, $\chi$,
scales with $L_s$ as  $\Sigma \sim L_s^{-\beta_m/\nu}$ 
and $\chi\sim L_s^{\gamma/\nu}$.

At each value of $\beta$ we made a linear $\chi^{2}$-fit for 
$\ln \Sigma$ and $\ln \chi$ versus $\ln L_{s}$.
The locations of the minima of $\chi^{2}/{\rm d.o.f.}$
as a function of $\beta$ provide estimates of $\beta_{c}$. 
We estimated the error in $\beta_{c}$ and the critical exponents 
by looking at the values of
$\beta$ that increase $\chi^{2}/{\rm d.o.f.}$ by one.
An estimate of the error in $\beta_{c}$ is  $\min(\chi^{2}/{\rm
d.o.f.})+1$, which also gives the error on the critical exponents. 
Fits at $L_{t}=6$ for the order
parameter and susceptibility gave 
$\beta_m/\nu=0.12(6)$, $\beta_{c}=0.7747(15)$ and 
$\gamma/\nu=1.66(9)$, $\beta_{c}=0.7750(15)$ respectively. Similar results
at $L_{t}=4$ are: $\beta_m/\nu=0.16(7)$, $\beta_{c}=0.6805(15)$, 
and $\gamma/\nu=1.57(9)$,  $\beta_{c}=0.6817(15)$. 

\begin{figure}[htb]

                \centerline{
 \psfig{silent=,file=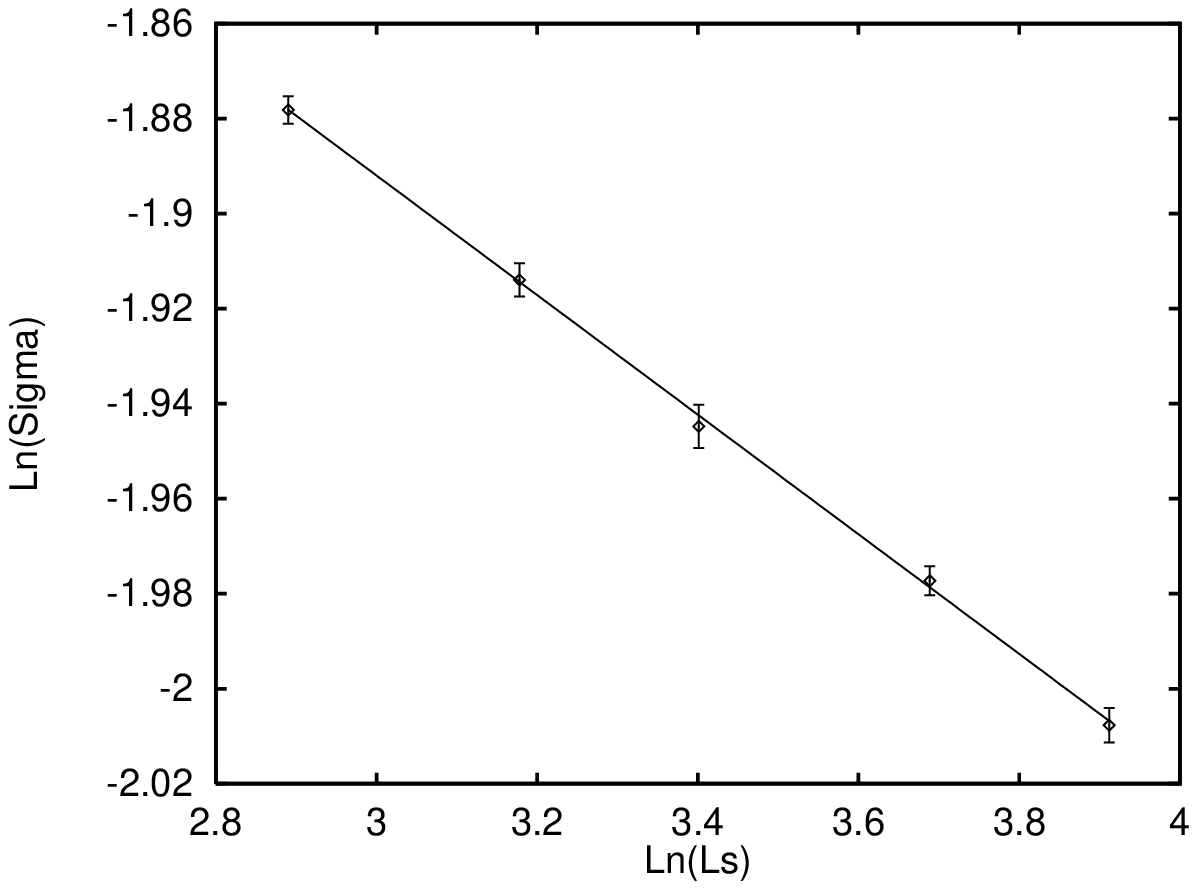,width=3.3in} 
 \psfig{silent=,file=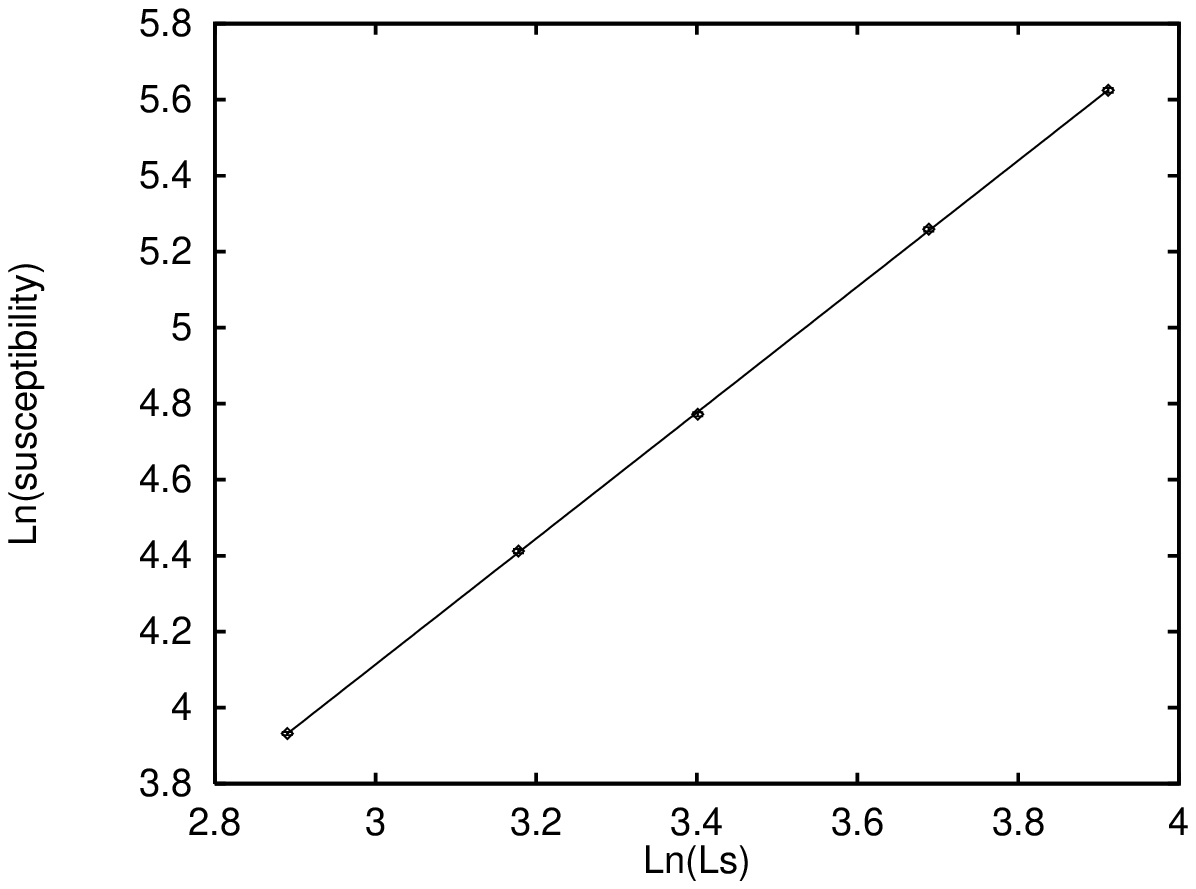,width=3.3in} }
\caption[]{Left: the best linear fit for $\ln \Sigma$ vs. $\ln L_{s}$ 
in the  minimum of $\chi^{2}$ ($\beta=0.7747$). Right: the same for
$\ln \chi$ vs. $\ln L_{s}$ ($\beta=0.7753$).}
\label{fig:sigma1}
\end{figure}

The critical exponents which we found are in a good 
agreement with the two-dimensional Ising exponents: 
$\beta_{m}/\nu=0.125$ and $\gamma/\nu=1.75$ (in the mean field
theory: $\beta_{m}/\nu=1$ and $\gamma/\nu=2$).
The values of $\beta_{c}$ extracted from this analysis 
are also consistent with the estimate of $\beta_{c}=0.7765(15)$
evaluated from the analysis of $V_{j}$.
Figs.~\ref{fig:sigma1} show 
the best linear fits of finite size dependence of $\ln\Sigma$ and 
$\ln\chi$ at $\beta=0.7747$ and $\beta=0.7753$ respectively.

\subsection{Scaling in the broken phase}

Our task in this section is to check the analytical 
result of eq.~(\ref{ginzburg}), which is equivalent to:
\begin{equation}\label{ginzburg-lat}
\xi \ll L_t N^x \ , \quad x=1/2.
\end{equation}
In other words, we want to determine the position of the crossover
from the mean-field (MF) critical behavior to the non-trivial
2d-Ising critical behavior as a function of $N$.

A straightforward way to do this is to study the dependence of the
order parameter, $\Sigma$, on $\beta$. Since $\Sigma$ vanishes at the
critical point, it can be thought of as a measure of the distance from
the criticality. We expect that for sufficiently small $\Sigma$, i.e.,
close to $\beta_c$, this dependence should be given by a power-law
scaling with the exponent $\beta_m=1/8$ of the 2d Ising model:
$\Sigma\sim({\rm const}-\beta)^{1/8}$.  This corresponds to the region
CD on the diagram Fig.~\ref{fig:phadi3}, i.e., $\xi \gg L_tN^x$.  For
larger $\Sigma$, further away from the criticality, the MF scaling
holds: $\Sigma\sim({\rm const}-\beta)^{1/2}$ (with some other value of
const). This is the region DB on Fig.~\ref{fig:phadi3}, i.e.,
$L_t\ll\xi\ll L_tN^x$. For even larger $\Sigma$ we should see the
scaling corresponding to the fixed point of the 3d Gross-Neveu model
\cite{HaKoKo93}: $\Sigma\sim({\rm const}-\beta)^1$. In this region
$1\ll\xi\ll L_t$.

Since we are interested in the boundary of the mean-field
scaling region we can find $\xi$ from $\Sigma$ using a well-known
relation between them: $\xi=1/(2\Sigma)$ \cite{NJL}, which holds
inside the mean-field region. Thus we avoid direct measurements of
$\xi$, which are much harder than measurements of $\Sigma$.

We studied the dependence of $\Sigma$ on $\beta$ at three
different values of $N=4,12,24$ on lattices with fixed $L_{t}=6$.
Ideally, in order to resolve all three critical scaling regions, or
windows, we need to provide: $1\ll L_t\ll L_t N^x \ll L_s$.  For
finite $L_t$, $L_s$, and $N_x$ these regions are squeezed, but for the
values we used one can clearly resolve the MF region with the
crossover towards the 2d-Ising region.

In contrast to the FSS analysis where $\xi\sim L_s$, now we need to keep
$\xi\ll L_s$ since we are studying bulk critical behavior.  We
monitored each simulation run for vacuum tunneling events, which
signal that $L_s$ is not big enough.  Away from the critical point
such events were so rare that good measurements of $\Sigma$ and its
susceptibility $\chi$ were possible. At couplings near the phase
transition we increased $L_s$ to suppress tunneling. Large
number $N$, such as $N=24$, suppresses fluctuations and allows an accurate
study within reasonable amount of computer resources. On the other hand, for
smaller $N$, such as $N=4$, the simulations were also very
efficient because the crossover to the 2d-Ising behavior starts
at a smaller correlation length.  The data from these simulations is
shown in Tables~\ref{tnf4},\ref{tnf12},\ref{tnf24}.

\begin{figure}[htb]

                \centerline{ \psfig{silent=,file=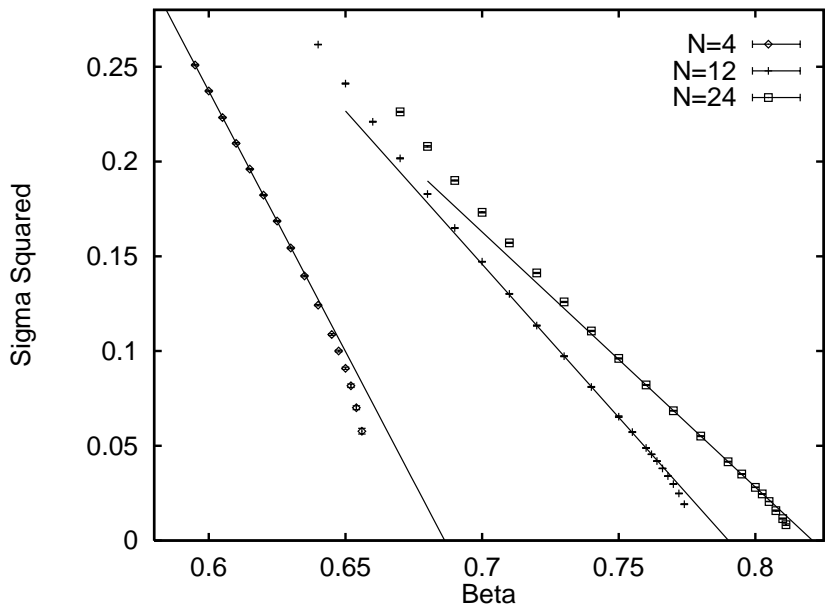,height=3.2in} }
\caption[]{Order parameter squared vs. $\beta$ for lattice theories
with $N=4,12,24$.
The straight lines are the fits to the data in the mean-field regions.}
\label{fig:broken}
\end{figure}

Since $\beta_{m}=1/2$ in the MF region, we fitted $\Sigma^{2}$ 
with a linear function of $\beta$. The linearity made the fitting procedure
very efficient. We evaluated the goodness of fit for data in various
ranges of $\beta$ in order to find the boundaries of the MF
region (see Tables~\ref{gnf4},\ref{gnf12},\ref{gnf24}).
The drop in the goodness of fit when new data are added implies that
the new points deviate from the MF behavior and they belong either to
the Gross-Neveu/MF crossover region or to the MF/2d-Ising 
crossover region.  Fig.~\ref{fig:broken} shows the data for
$\Sigma^{2}$ versus $\beta$, where the straight lines
represent the best linear fits in the MF region. 
All three graphs show the MF/2d-Ising crossovers.  
The graphs for $N=12,24$ also clearly show the Gross-Neveu/MF crossover.

\begin{figure}[hbt]

                \centerline{ \psfig{silent=,file=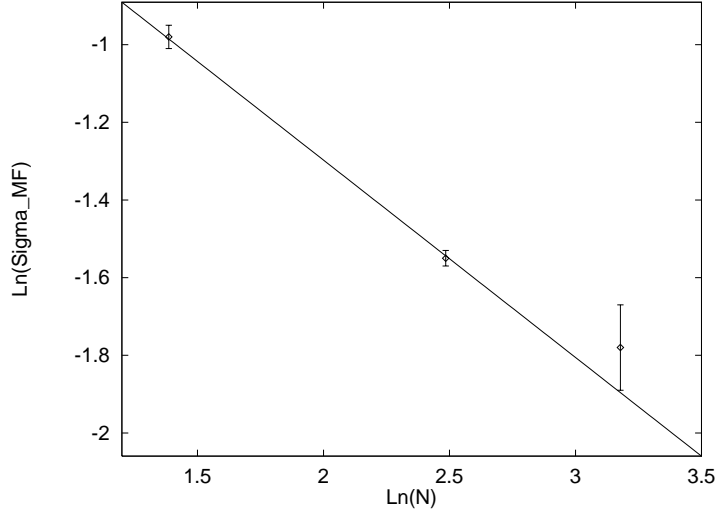,height=2.7in} }
\caption[]{The best power-law fit (goodness 0.26) 
for the dependence of the boundary of the mean-field region, 
$\Sigma_{\rm MF}$, on the number of fermions, $N$.}
\label{fig:power}
\end{figure}

We used the histogram reweighting method
in order to extract the value of $\Sigma$ at the ``border'' of the MF
region with the 2d-Ising region. The quality of the fit
drops very sharply. This allowed us to use a very conservative
estimate of the error, such as the width of the region of $\Sigma$
in which the quality of the fit drops down to 1\% (from, e.g., 90\%
in the case of $N=4$). The center of this region of the drop is taken as
the boundary of the MF scaling window. We find $\Sigma_{\rm MF}=0.377(11)$,
for $N=4$, $\Sigma_{\rm MF}=0.213(4)$ for $N=12$, and 
$\Sigma_{\rm MF}=0.168(20)$ for $N=24$.
It is clear that the non-trivial 2d-Ising region is squeezed as $N$
increases.
In order to check the analytical prediction of eq.~(\ref{ginzburg}),
or (\ref{ginzburg-lat}), we fitted our results
to the form $\Sigma_{\rm MF}={\rm const}\cdot N^{-x}$ 
(see Fig.~\ref{fig:power}). 
We found $x=0.51(3)$ which is in agreement
with the analytical prediction $x=0.5$.

\section{Discussion and conclusions}

In this paper we demonstrated, using analytical arguments and
Monte Carlo simulations, that the width of the region near
the finite temperature chiral phase transition where a 
non-trivial critical behavior sets in is suppressed in theories
with large number of fermions, $N$. This phenomenon explains
an apparent contradiction between the large-$N$ expansion
\cite{RoSp94,KoKo94}, predicting mean-field scaling, and more
general arguments based on dimensional reduction and universality
\cite{PiWi84}, predicting the non-trivial scaling of a scalar
theory in $d-1$ dimensions. 

A tentative explanation of \cite{RoSp94} was based, somewhat implicitly,
on the expectation that the large-$N$ expansion breaks down and
cannot predict correct non-trivial exponents. Our key
point is that we can say {\em when}, and as a function
of what parameter (i.e., as a function of the distance from the
criticality, $m/T$) 
the large-$N$ expansion breaks down. What is, perhaps, even
more important, is that we show that there does exist a region 
where the large-$N$ expansion {\em is} valid. This region (in the
space of $m/T$) of the mean-field behavior squeezes out the region
of the true non-trivial critical behavior.%
\footnote{ It was conjectured in \cite{Ra95} that the Ginzburg region
of non-trivial scaling could turn out to be small.}

One can also look at the whole problem as a question of the order of
limits.  If the limit $N\to\infty$ is taken before $m/T\to0$, the
non-trivial region disappears, and all critical behavior is given by
mean-field theory. If, on the other hand, the limit $m/T\to0$ is taken
at fixed $N$, the true non-trivial scaling will hold. These effects
can be clearly seen in our Monte Carlo studies.

A helpful analogy can be drawn with the question of the long distance
behavior of the scalar field correlator in the original 2-dimensional
Gross-Neveu model with a $U(1)$ symmetry at large $N$ \cite{GN}.  The
large-$N$ expansion predicts spontaneous symmetry breaking and
long-range order, which is in clear contradiction with the
Mermin-Wagner theorem.  As was shown by Witten \cite{Wi78}, there is
an ``almost long-range order'' in the system, i.e., the correlator
falls off like $r^{-1/N}$. Looking at this expression, one can
easily see the interplay between the limits $r\to\infty$ and
$N\to\infty$.
 
We applied our arguments to a specific
example of a Yukawa theory, but the mechanism responsible for
this phenomenon is clearly more universal and may apply, in
particular, to QCD, provided that the number of colors $N_c$
is large. The Yukawa, or the Nambu-Jona-Lasinio, theory
is well-known to provide a very good description for the
phenomenon of chiral symmetry breaking and restoration. 
One can then think of the theory we considered as an
effective description of the degrees of freedom participating
in chiral symmetry breaking. 

As we have seen, the role of the fermions is to screen the effective
self-coupling of the scalar field, $\lambda$. The strength of this
effect depends on two factors: (i) large $N$, and (ii) large window of
scales between the cutoff of the effective theory, $\Lambda$, and the
temperature, $T$.  Let us see if these conditions are fulfilled in
QCD. The scale of the spontaneous symmetry breaking is of order
$\Lambda\sim 1$ GeV, while $T_c\approx 160$ MeV.  Thus, there is
almost an order of magnitude window between $\Lambda$ and $T$, which
is presumably sufficient to drive the effective self-coupling of the
scalar field to its infrared fixed point at the scale of $T_c$. How
small this value is now depends on the number of the fermions (the
condition (i)). In QCD the value $N_c=3$, though not very large, can
be considered large in some cases. It would be interesting to see if
this phenomenon could be rigorously shown to occur in the limit
$N_c\to\infty$, which is very plausible.

The effect of the suppression of the width of the non-trivial critical
region in QCD may lead to the following prediction: only mean-field
(but no non-trivial $O(4)$) scaling behavior could be seen
because of non-zero quark masses, $m_q$.  On the one hand, the
mean-field scaling would hold until relatively large thermal
correlation length.  On the other hand, this correlation length in the
real world is limited by the quark masses, $m_q$. 

The following crude estimates can serve to illustrate this point.%
\footnote{To make these estimates quantitative
one needs to find numerical coefficients, which are determined by
non-universal dynamical properties of QCD and by the definition
of the boundary of the mean-field region.} 
A true criticality is never reached near $T_c$, because
$m_q$ plays the same role as the external ordering magnetic field in a
ferromagnet.  The largest thermal correlation length in units of
$1/T$, $T/m$, can be estimated using the analogy to a ferromagnet
and the mean-field value of $\nu/(\beta\delta)=1/3$, as%
\footnote{We use the definition of the exponents: $m\sim t^\nu$,
$\Sigma\sim t^\beta$, and $\Sigma\sim m_q^{1/\delta}$, together
with the central postulate of the scaling theory:
critical properties are determined by the correlation
length, $1/m$.
}%
:
\begin{equation}
{T\over m} \sim \left(T_c\over m_q\right)^{\nu\over\beta\delta}\approx 
\left(160{\rm\ MeV}\over 5 {\rm\ MeV}\right)^{1/3}\sim 3.
\label{qcd}
\end{equation}
The fact that this number is not large could be guessed by observing
that the zero temperature pion masses, which are driven by $m_q$, are
as large as $T_c$. This largest correlation length may turn out to be
smaller than the one required for the crossover to non-trivial scaling
region: $T/m\sim N_c\ln(\Lambda/T_c)\sim 6$, according to
(\ref{ginzburg,d=4}). In this case, the (near-)critical behavior observed
in the window allowed by non-zero quark masses (according to
(\ref{qcd}), roughly: $1<T/m<3$) will be given entirely by the 
mean-field scaling.

\subsection*{Acknowledgments}

Discussions with A. Kocic, R. Pisarski and T. Tran are greatly appreciated.
We learned that a result similar to (\ref{ginzburg}) had been
independently derived by Pisarski \cite{Pi}. This work was supported
in part by the NSF grants PHY96-05199 and PHY97-22101.

\begin{table}[p]
\caption[]{Simulations for the FSS analysis with $L_{t}=6$ and $N=12$.}
\begin{center}
\parbox{3in}{
\begin{tabular}{ccc} 
$L_{s}$  &  $\beta_{\rm sim}$  &  trajectories\\
\hline\hline
18    &       0.7744       &          130,000 \\
18    &       0.7763       &          100,000 \\
24    &       0.7744       &          100,000 \\
24    &       0.7764       &           80,000 \\
\hline
\end{tabular}
}
\parbox{3in}{
\begin{tabular}{ccc} 
$L_{s}$  &  $\beta_{\rm sim}$  &  trajectories\\
\hline\hline
30    &       0.7744       &          120,000 \\
30    &       0.7764       &          130,000 \\
40    &       0.7754       &          140,000 \\
50    &       0.7754       &          152,000 \\
\hline
\end{tabular}
}
\end{center}
\label{tfss}
\end{table}

\begin{table}[p]
\caption[]{The data for the scaling of $\Sigma$ in the broken phase. $N=4$.}
\bigskip
\parbox{3in}{
\begin{tabular}{cccc} 
$\beta$ &  $\Sigma$ &  $L_{s}$ & trajectories\\
\hline\hline
0.595   &  0.5008(4) &   36    & 30,000 \\
0.600   &  0.4871(4) &   36    & 40,000 \\
0.605   &  0.4724(4) &   36    & 40,000 \\
0.610   &  0.4578(4) &   36    & 40,000 \\
0.615   &  0.4426(4) &   36    & 50,000 \\
0.620   &  0.4268(4) &   36    & 54,000 \\
0.625   &  0.4106(3) &   36    & 70,000 \\
0.630   &  0.3928(4) &   36    & 64,000 \\
\hline
\end{tabular}
}
\parbox{3in}{
\begin{tabular}{cccc} 
$\beta$ &  $\Sigma$ &  $L_{s}$ & trajectories\\
\hline\hline
0.635   &  0.3736(7) &   36    & 70,000 \\
0.640   &  0.3524(5) &   36    & 110,000 \\
0.645   &  0.3297(8) &   36    & 130,000 \\
0.6475  &  0.3162(9) &   36    & 150,000 \\
0.650   &  0.3016(10) &  48    & 54,000 \\
0.652   &  0.2858(15) &  60    & 41,000 \\
0.654   &  0.2647(26) &  80    & 25,400 \\
0.656   &  0.2398(30) &  80    & 30,000 \\
\hline
\end{tabular}
}
\label{tnf4}
\end{table}

\begin{table}[p]
\caption[]{The data for the scaling of $\Sigma$ in the broken phase. $N=12$.}
\bigskip
\parbox{2in}{
\begin{tabular}{ccc}
$\beta$ &  $\Sigma$ &  $L_{s}$ \\
\hline\hline
0.640  &  0.5112(2)  &  30 \\
0.650  &  0.4911(3)  &  30 \\
0.660  &  0.4701(2)  &  30 \\
0.670  &  0.4490(3)  &  30 \\
0.680  &  0.4276(2)  &  30 \\
0.690  &  0.4061(3)  &  30 \\
0.700  &  0.3836(2)  &  30 \\
\hline
\end{tabular}
}
\parbox{2in}{
\begin{tabular}{ccc}
$\beta$ &  $\Sigma$ &  $L_{s}$ \\
\hline\hline
0.710  &  0.3608(4)  &  30 \\
0.720  &  0.3368(4)  &  30 \\
0.730  &  0.3119(4)  &  30 \\
0.740  &  0.2847(4)  &  30 \\
0.750  &  0.2557(8)  &  30 \\
0.755  &  0.2393(5)  &  30 \\
0.760  &  0.2199(5)  &  30 \\
\hline
\end{tabular}
}
\parbox{2in}{
\begin{tabular}{ccc}
$\beta$ &  $\Sigma$ &  $L_{s}$ \\
\hline\hline
0.764  &  0.2048(2)  &  60 \\
0.766  &  0.1951(3)  &  60 \\
0.768  &  0.1844(3)  &  60 \\
0.770  &  0.1725(3)  &  60 \\
0.772  &  0.1575(5)  &  60 \\
0.774  &  0.1388(7)  &  60 \\
&&\\
\hline
\end{tabular}
}
\label{tnf12}
\end{table}

\begin{table}[p]
\caption[]{The data for the scaling of $\Sigma$ in the broken phase. $N=24$.}
\bigskip
\parbox{3in}{
\begin{tabular}{cccc}
$\beta$ &  $\Sigma$ &  $L_{s}$ & trajectories \\
\hline\hline
0.6700  &  0.4756(4)  &  24  &  3,000 \\
0.6800  &  0.4560(4)  &  24  &  3,000 \\
0.6900  &  0.4359(4)  &  24  &  3,000 \\
0.7000  &  0.4162(4)  &  24  &  3,000 \\
0.7100  &  0.3963(4)  &  24  &  3,000 \\
0.7200  &  0.3760(5)  &  24  &  3,000 \\
0.7300  &  0.3551(5)  &  24  &  3,000 \\
0.7400  &  0.3325(5)  &  24  &  3,000 \\
0.7500  &  0.3101(5)  &  24  &  3,000 \\
0.7600  &  0.2866(3)  &  24  &  8,000 \\
\hline
\end{tabular}
}
\parbox{3in}{
\begin{tabular}{cccc}
$\beta$ &  $\Sigma$ &  $L_{s}$ & trajectories \\
\hline\hline
0.7700  &  0.2617(4)  &  24  &  8,000 \\
0.7800  &  0.2345(5)  &  24  &  8,000 \\
0.7900  &  0.2035(5)  &  24  &  10,000 \\
0.7950 &  0.1874(3)  &  36  &  15,000 \\
0.8000 &  0.1676(3)  &  36  &  20,000 \\
0.8025 & 0.1568(4)  &  36  &  20,000 \\
0.8050 & 0.1437(4)  &  48  &  24,000 \\
0.8075 & 0.1267(10) & 60  &  12,000 \\
0.8100 & 0.1074(15) & 60  &  22,000 \\
0.8112 & 0.0914(28) & 72  &  10,000 \\
\hline
\end{tabular}
}
\label{tnf24}
\end{table}

\begin{table}[p]
\caption[]{Goodness of linear fits of $\Sigma^2$ vs. $\beta$ for
various ranges of $\beta$. $N=4$.}
\bigskip
\parbox{3in}{
\begin{tabular}{ccc} 
no. of points  & $\beta$ range  &  goodness\\
\hline\hline
4   &  0.595 - 0.610   &  0.92 \\
5   &  0.595 - 0.615   &  0.95 \\
6   &  0.595 - 0.620   &  0.98 \\
7   &  0.595 - 0.625   &  0.99 \\
\hline
\end{tabular}
}
\parbox{3in}{
\begin{tabular}{ccc} 
no. of points  & $\beta$ range  &  goodness\\
\hline\hline
8   &  0.595 - 0.630   &  0.96 \\
9   &  0.595 - 0.635   &  0.21 \\
10  &  0.595 - 0.640   &  $10^{-5}$ \\
11  &  0.595 - 0.645   &  $10^{-11}$ \\
\hline
\end{tabular}
}
\label{gnf4}
\end{table}

\begin{table}[p]
\caption[]{The same as Table \ref{gnf4} but for $N=12$.}
\bigskip
\parbox{3in}{
\begin{tabular}{ccc} 
no. of points  & $\beta$ range  &  goodness\\
\hline\hline
5   &  0.70 - 0.74  &  $4\times 10^{-3}$ \\
6   &  0.70 - 0.75  &  $5\times 10^{-3}$ \\
5   &  0.71 - 0.75  &  0.20 \\
4   &  0.72 - 0.75  &  0.67 \\
5   &  0.72 - 0.755 &  0.77 \\
\hline
\end{tabular}
}
\parbox{3in}{
\begin{tabular}{ccc} 
no. of points  & $\beta$ range  &  goodness\\
\hline\hline
6   &  0.72 - 0.760 &  0.60 \\
7   &  0.72 - 0.762 &  0.40 \\
8   &  0.72 - 0.764 &  $5\times 10^{-4}$ \\
9  &  0.72 - 0.766 &  $2\times 10^{-15}$ \\
&&\\
\hline
\end{tabular}
}
\label{gnf12}
\end{table}

\begin{table}[p]
\caption[]{The same as Table \ref{gnf4} but for $N=24$.}
\bigskip
\parbox{3.2in}{
\begin{tabular}{ccc} 
no. of points  & $\beta$ range  &  goodness\\
\hline\hline
7  &  0.74 - 0.795   &  $6\times 10^{-4}$ \\
8  &  0.74 - 0.800   &  $1.2\times 10^{-3}$ \\
6  &  0.75 - 0.795   &  0.28 \\
7  &  0.75 - 0.800   &  0.24 \\
3  &  0.76 - 0.780   &  0.68 \\ 
4  &  0.76 - 0.790   &  0.91 \\ 
\hline
\end{tabular}
}
\parbox{3.2in}{
\begin{tabular}{ccc} 
no. of points  & $\beta$ range  &  goodness\\
\hline\hline
5  &  0.76 - 0.795   &  0.70 \\ 
6  &  0.76 - 0.800   &  0.38 \\
7  &  0.76 - 0.8025  &  0.17 \\
8  &  0.76 - 0.8050  &  $1.2\times 10^{-8}$ \\  
9  &  0.76 - 0.8075  &  $5\times 10^{-16}$ \\
&&\\
\hline
\end{tabular}
}
\label{gnf24}
\end{table}

\end{document}